\documentclass[aps,twocolumn,showpacs,preprintnumbers,amsmath,amssymb,superscriptaddress,nofootinbib]
{revtex4}

\usepackage{graphicx}
\usepackage{dcolumn}
\usepackage{bm}
\usepackage{hyperref}

\def\mk{\mathbf{k}}

\def\detbeta{\mathrm{det}\widetilde{\beta}}
\def\Deltabeta{\Delta\widetilde{\beta}}
\def\Omegaxy{\widetilde{\Omega}_{xy}^{(+)}}
\def\Omegayx{\widetilde{\Omega}_{yx}^{(+)}}
\def\MaxOmegaxy{\widetilde{\Omega}_{xy}^{(+,P)}}
\def\MaxOmegayx{\widetilde{\Omega}_{yx}^{(+,P)}}

\newcommand{\secref}[1]{Sec.~\ref{#1}}
\newcommand{\figref}[1]{Fig.~\ref{#1}}

\begin{document}

\title
{Maximum intrinsic spin-Hall conductivity in two-dimensional
systems with k-linear spin-orbit interaction}

\author{Tsung-Wei Chen}
\email{twchen@mail.nsysu.edu.tw}\affiliation{Department of
Physics, National Sun Yat-sen University, Kaohsiung 80424, Taiwan}

\date{\today}

\begin{abstract}
We analytically calculate the intrinsic spin-Hall conductivity
(ISHC) ($\sigma^z_{xy}$ and $\sigma^z_{yx}$) in a clean,
two-dimensional system with generic k-linear spin-orbit
interaction. The coefficients of the product of the momentum and
spin components form a spin-orbit matrix $\widetilde{\beta}$. We
find that the determinant of the spin-orbit matrix $\detbeta$
describes the effective coupling of the spin $s_z$ and orbital
motion $L_z$. The decoupling of spin and orbital motion results in
a sign change of the ISHC and the band-overlapping phenomenon.
Furthermore, we show that the ISHC is in general unsymmetrical
($\sigma^z_{xy}\neq-\sigma^z_{yx}$), and it is governed by the
asymmetric response function $\Deltabeta$, which is the difference
in band-splitting along two directions: those of the applied
electric field and the spin-Hall current. The obtained
non-vanishing asymmetric response function also implies that the
ISHC can be larger than $e/8\pi$, but has an upper bound value of
$e/4\pi$. We will that the unsymmetrical properties of the ISHC
can also be deduced from the manifestation of the Berry curvature
at the nearly degenerate area. On the other hand, by investigating
the equilibrium spin current, we find that $\detbeta$ determines
the field strength of the SU(2) non-Abelian gauge field.

\end{abstract}
\pacs{71.70.Ej, 72.25.Dc, 73.63.Hs} \maketitle

\section{Introduction}
Within condensed matter physics, spintronics has in itself become
a strong field for considerable research, owing to not only its
potential applications in electronic technologies but also the
many fundamental questions that are raised on the physics of
electron spin.~\cite{pri98} Particularly, the spin-orbit
interaction recently has strongly attracted the attentions of
theoreticians and experimenters since it opens up the possibility
of manipulating electron (or hole) spin in non-magnetic materials
by electrical means.~\cite{zut04,Hisch99} Since the theoretical
prediction of the spin-Hall effect, the application of spintronics
has seen considerable advancement. It was shown that the Mott-type
skew scattering by impurities would result in separation of
opposite spin states via the spin-orbit interaction to the
impurity atom.~\cite{Hisch99} This is the extrinsic spin-Hall
effect. Nevertheless, it has been found that in p-type
\cite{Murakami03} (Luttinger model) and n-type \cite{Sinova04}
(Rashba model) semiconductors, the spin-polarized current
(electron or hole) can be generated by the intrinsic spin-orbit
interaction in non-magnetic structure in the absence of impurity
scattering, which is called the intrinsic spin-Hall effect (ISHC).

The calculation of spin-Hall conductivity (SHC) plays a crucial
role in studying the spin-Hall effect as it can be in comparison
with experimental result. The extrinsic spin-Hall effect was
experimentally discovered in the three-dimensional (3D) n-type
GaAs by optical means via spin accumulation at the edges of a
sample~\cite{Kato04,Stern06} and in two-dimensional (2D) n-type
AlGaAs/GaAs.~\cite{Sih05} The magnitude of the experimental value
of SHC [$0.5 (1/em \Omega)$] in Ref.~\cite{Kato04} agrees with its
theoretical value [$0.9 (1/em\Omega)$].~\cite{Engel05} However,
the sign of the theoretical SHC is opposite to the experimental
value, and it needs to be further clarified.~\cite{Engel05} In 2D
p-type AlGaAs/GaAs~\cite{Wund05}, the experimental value of SHC
[$2.5(e/8\pi)$] also agrees with the theoretical value
[$1.9(e/8\pi)$] in order of magnitude.~\cite{Berne05}.
Particularly, in Ref.~\cite{Berne05}, the clean limit is
considered in the calculation. In 2D n-type InGaN/GaN, the
strain-dependent intrinsic spin-Hall effect detected by optical
means is explained in terms of SHC in which the strain effect is
included.~\cite{Chang06} In 3D metal Pt wire, the large ISHC
measured electrically throughout the inverse spin-Hall effect at
room temperature is $240(\hbar/e\Omega
cm)$.~\cite{Saitoh06,Kimura07} It was theoretically explained in
Ref.~\cite{Guo08} on the basis of the huge Berry
curvature~\cite{boh03} near the Fermi level at the L and X
symmetry point in the Pt Brillouin zone; the obtained theoretical
value of ISHC is $200(\hbar/e\Omega cm)$ in the absence of
impurity scattering. Most recently, a large spin-Hall signal is
observed at room temperature in FePt/Au multi-terminal
devices.~\cite{Seki08}

Importantly, both the direction of the applied electric field and
the strength of the spin-orbit interactions alters the values of
the ISHC. For the former case, a typical example is the
Rashba-Dresselhaus system.~\cite{RD} When an electric field is
applied along the $x$ ($[010]$) (or $y$ $[100]$) direction, we
obtain $\sigma^z_{xy}=-\sigma^z_{yx}$, and these values are equal
to the universal constant $e/8\pi$. However, if an electric field
is applied along $x'$ ($[110]$) (or $y'$, i.e., $[1\bar{1}0]$), we
obtain $\sigma^z_{x'y'}\neq-\sigma^z_{y'x'}$, and one of these
values has a value higher than $e/8\pi$. The later case requires a
systematical investigation because the spin-orbit interaction
could be very complicated. For example, it has been proposed that
a strained semiconductor results in various k-linear
band-splitting.~\cite{BernePRB05} Nevertheless, we find that
strain-induced spin splitting together with the spin-orbit
coupling of the host semiconductor can be simplified and expressed
in terms of the coefficients of the spin-orbit matrix [see Eq.
(\ref{Hamiltonian-beta})]. In this study, we focus on generic 2D
k-linear spin-orbit coupled systems without impurity scattering,
and we systematically investigate the effects of spin-orbit
interactions and the direction of the applied electric field on
spin-Hall current. We find that the ISHC can be calculated
analytically and that its unsymmetrical properties can be
described using a unified approach.

We show that $\detbeta$ [see Eq. (\ref{detbeta})] is expressed as
the effective coupling of the z-component of spin $s_z$, and
orbital angular momentum $L_z$. The decoupling of spin and orbital
motion associated with the band-overlapping phenomenon results in
the vanishing and sign change of the ISHC.

Furthermore, by analytically calculating the ISHC, we find that
the unsymmetrical result of the ISHC
($\sigma_{xy}^z\neq\sigma_{yx}^z$) is governed by the asymmetric
response function $\Deltabeta$ [see Eq. (\ref{Deltabeta})], which
is the difference in band-splitting in two directions: those of
the applied electric field and the spin-Hall current. We find that
the direction of the applied electric field alters the magnitude
of the asymmetric response function. Consequently, we show that
there exists a specific direction of applied electric field such
that the asymmetric response function reaches a maximum value. In
this case, we show that the ISHC also reaches a maximum value in
the range $e/8\pi$ to $e/4\pi$, where $e/4\pi$ is the upper bound
value of the ISHC. The unsymmetrical result and the maximum
asymmetric response function of the ISHC can also be deduced from
the behavior of the Berry curvature at the nearly degenerate area.
The nearly degenerate area refers to the area where the inner and
outer bands are very close to each other on the Fermi surface.

Our present paper is organized as follows. In \secref{sec:ISHC},
we define the spin-orbit matrix obtained from the coefficients of
the product of momentum and spin. The intrinsic spin-Hall
conductivity is shown to be proportional to the determinant of
spin-orbit matrix $\detbeta$. We use Foldy-Wouthuysen
transformation to show that the effective coupling of the spin
$z$-component $s_z$ and orbital angular momentum $L_z$ is
$-2m(\detbeta)/\hbar^4$. In \secref{sec:MVSHC}, we analytically
calculate the ISHC of the generic 2D k-linear spin-orbit coupled
system. The asymmetric response function and the upper bound value
of ISHC will be discussed. In \secref{sec:A-ISHC}, in order to
reveal the maximum value of ISHC, the direction of applied
electric field and its influence on the asymmetric response
function is studied. In \secref{sec:BerryC}, we will show that the
unsymmetrical properties of ISHC can be deduced from the variation
of Berry curvature. In \secref{sec:ESC}, we discuss the
relationship between equilibrium spin current and spin-orbit
matrix. We show that $\detbeta$ plays the role of color magnetic
field strength. Our conclusions are presented in \secref{sec:con}.

\section{Intrinsic Spin-Hall conductivity}\label{sec:ISHC}
\subsection{spin-orbit matrix and ISHC}
The 2D k-linear spin-orbit coupled system Hamiltonian in the
presence of an applied electric field can be written as
\begin{equation}\label{Hamiltonian}
H=\varepsilon_{\mk}+H_{so}+V(x,y),\\
\end{equation}
where
\begin{equation}\label{Hamiltonian-beta}
\begin{split}
H_{so}&=\sum_{ij}\beta_{ij}\sigma_ik_j\\
&=\left(\sigma_x~~\sigma_y\right)\left(\begin{array}{cc}
\beta_{xx}&\beta_{xy}\\
\beta_{yx}&\beta_{yy}
\end{array}\right)\left(\begin{array}{c}
k_x\\
k_y
\end{array}\right).
\end{split}
\end{equation}
The kinetic energy is $\varepsilon_{\mk}=\hbar^2k^2/2m$ and
$\sigma_i~(i=x,y)$ are the Pauli spin matrices. The external
potential $V(x,y)$ is $V(x,y)=e\mathbf{E}\cdot\mathbf{x}$. The
generic k-linear spin-orbit coupled 2D systems are related to the
spin-orbit matrix $\widetilde{\beta}$,
\begin{equation}
\widetilde{\beta}=\left(\begin{array}{cc}
\beta_{xx}&\beta_{xy}\\
\beta_{yx}&\beta_{yy}
\end{array}\right),
\end{equation}
where the coefficients $\beta_{ij}$ represent the spin-orbit
interactions in 2D systems. As an example, consider the Rashba
system [$\alpha(\sigma_xk_y-\sigma_yk_x)$]~\cite{BRashba}, the
pure Dresselhaus system
[$-\beta(\sigma_xk_x-\sigma_yk_y)$]~\cite{Dre55} and the
Dirac-type system
[$g(\sigma_xk_x+\sigma_yk_y)$]~\cite{McClure56,Sin07}; the
corresponding spin-orbit matrices for these systems are
\begin{equation}
\widetilde{\beta}_R=\alpha\left(\begin{array}{cc}
0&1\\
-1&0
\end{array}\right),~\widetilde{\beta}_S=\beta\left(\begin{array}{cc}
-1&0\\
0&1
\end{array}\right),~\widetilde{\beta}_D=g\left(\begin{array}{cc}
1&0\\
0&1
\end{array}\right),
\end{equation}
respectively. Another example is the spin splitting in a bulk
strained semiconductor.~\cite{BernePRB05,Pikus84} The spin-orbit
matrices $\widetilde{\beta}_1$ and $\widetilde{\beta}_2$ denote,
respectively, the system with structure inversion asymmetry (SIA)
strain-induced splitting and the system with bulk inversion
asymmetry (BIA) strain-induced splitting. They are given by
\begin{equation}
\begin{split}
&\widetilde{\beta}_1=\frac{1}{2}C_3\epsilon_{xy}\left(\begin{array}{cc}
0&1\\
0&0
\end{array}\right)-\frac{1}{2}C_3\epsilon_{yx}\left(\begin{array}{cc}
0&0\\
1&0
\end{array}\right)\\
&\widetilde{\beta}_2=D(\epsilon_{zz}-\epsilon_{yy})\left(\begin{array}{cc}
1&0\\
0&0
\end{array}\right)+D(\epsilon_{xx}-\epsilon_{zz})\left(\begin{array}{cc}
0&0\\
0&1
\end{array}\right),
\end{split}
\end{equation}
where the structure constant $C_3>0$ and
$D>0$.~\cite{BernePRB05,kzterm} Thus, in addition to SIA and
bulk-inversion-symmetry breaking induced spin-orbit interaction,
the strain-induced spin splitting is included in the spin-orbit
matrix elements. Accordingly, we do not pose any restrictions on
the spin-orbit matrix elements in the following calculations. For
calculating ISHC, we further rewrite Eq. (\ref{Hamiltonian}) in
the following form.
\begin{equation}
H=\varepsilon_{\mk}+d_x\sigma_x+d_y\sigma_y,
\end{equation}
where
\begin{equation}\label{dxdy}
\begin{split}
&d_x=\beta_{xx}k_x+\beta_{xy}k_y\\
&d_y=\beta_{yx}k_x+\beta_{yy}k_y.
\end{split}
\end{equation}
The eigenenergy is $E_{n\mk}=\varepsilon_{\mk}-nd$ for two
branches $n=\pm$ ($n=+$ for outer band and $n=-$ for inner band),
where the dispersion term $d=\sqrt{d_x^2+d_y^2}$ can be written as
\begin{equation}\label{dis1}
d=k\Gamma(\phi),
\end{equation}
where
\begin{equation}\label{dis2}
\begin{split}
\Gamma(\phi)^2&=\left(\beta_{xx}^2+\beta_{yx}^2\right)\cos^2\phi+\left(\beta_{xy}^2+\beta_{yy}^2\right)\sin^2\phi\\
&+\left(\beta_{xx}\beta_{xy}+\beta_{yx}\beta_{yy}\right)\sin(2\phi).
\end{split}
\end{equation}
The energy dispersion Eq. (\ref{dis2}) satisfies
$\Gamma(\phi)=\Gamma(\phi+\pi)$ because the time-reversal symmetry
is preserved. For a positive chemical potential ($\mu>0$), the
Fermi momenta for two branches satisfy the following condition
\begin{equation}\label{diff-FM}
k_F^{+}-k_F^{-}=\frac{2m\Gamma(\phi)}{\hbar^2},
\end{equation}
which is the band-splitting at $\phi$ direction on the Fermi
surface. The ISHC can be evaluated by using the Kubo
formula~\cite{Mahan}
\begin{equation}
\sigma^{z}_{ij}(\omega)=\lim_{\omega\rightarrow0}\frac{Q_{ij}^{z}(\omega)}{-i\omega}.
\end{equation}
$Q_{ij}^{z}(\omega)$ is the spin current-charge current
correlation function. The index $j$ represents the direction of
applied electric field and $i$ the direction of response current.
The conventional definition of spin current is
$J_{i}^{z}=\frac{\hbar}{4}\left\{\partial H/\partial
k_i,\sigma_z\right\}$~\cite{Shi06}, and charge current is defined
as $J_j=e\partial H/\partial\hbar k_j$. From the standard
approach, it can be shown that~\cite{BernePRB05-2}
\begin{equation}
Q^{z}_{ij}(\omega)=i\omega\frac{e}{V}\sum_{\mk}\frac{f_{\mk+}-f_{\mk-}}{d(\omega^2-4d^2)}\frac{\partial\varepsilon_{\mk}}{\partial
k_i}\left(d_x\frac{\partial d_y}{\partial k_j}-d_y\frac{\partial
d_x}{\partial k_j}\right),
\end{equation}
where $f_{\mk\pm}$ represents the Fermi function for two energy
branches. Note that the correlation function contains the kinetic
term. Next, we focus on spin-Hall conductivity in the static case
($\omega=0$). When an electric field is applied in $k_y$
direction, and the spin-Hall response in $k_x$ direction is given
by
\begin{equation}\label{SHC}
\sigma^{z}_{xy}=-\frac{e}{V}\sum_{\mk}\frac{f_{\mk+}-f_{\mk-}}{-4d^3}\frac{\partial\varepsilon_{\mk}}{\partial
k_x}\left(d_x\frac{\partial d_y}{\partial k_y}-d_y\frac{\partial
d_x}{\partial k_y}\right).
\end{equation}
Substituting Eqs. (\ref{dxdy}), (\ref{dis1}), (\ref{dis2}), and
(\ref{diff-FM}) into Eq.(\ref{SHC}) and using the replacement
$(1/V)\sum_{\mk}\rightarrow\int kdkd\phi/(2\pi)^2$, after a
straightforward calculation, we obtain
\begin{equation}\label{SHC-1}
\sigma^{z}_{xy}=\frac{e}{8\pi^2}\left(\detbeta\right)\int_0^{2\pi}d\phi\frac{\cos^2\phi}{\Gamma(\phi)^2},
\end{equation}
where $\detbeta$ stands for the determinant of the spin-orbit
matrix $\widetilde{\beta}$
\begin{equation}\label{detbeta}
\detbeta=\beta_{xx}\beta_{yy}-\beta_{xy}\beta_{yx}.
\end{equation}
Equation. (\ref{SHC-1}) indicates that the spin-Hall conductivity
vanishes when $\detbeta=0$.  To understand the vanishing ISHC, we
have to refer to the effective coupling of spin $s_z$ and orbital
motion $L_z$ in the presence of an applied electric field.  In the
following subsection, we will show that the effective coupling of
orbital motion and spin is related to Eq. (\ref{detbeta}).

\subsection{effective coupling of spin and orbital motion}
We can apply the Foldy-Wouthuysen transformation~\cite{Foldy} to
the Hamiltonian Eq. (\ref{Hamiltonian}), and diagonalize the
Hamiltonian up to some order of $\beta_{ij}$. Because $\detbeta$
is order of $\beta^2_{ij}$, the unitary transformation that can
diagonalize Eq. (\ref{Hamiltonian}) up to second order is given by
(see Appendix~\ref{Appendix:FW})
\begin{equation}\label{Unitary}
\begin{split}
&U=\exp\left\{-i\frac{m}{\hbar^2}D\right\},\\
&D=\sigma_xF_x+\sigma_yF_y,\\
&F_x=\beta_{xx}x+\beta_{xy}y,\\
&F_y=\beta_{yx}x+\beta_{yy}y,
\end{split}
\end{equation}
where $F_x$ and $F_y$ are obtained by using the replacements
$k_x\rightarrow x$ and $k_y\rightarrow y$ in $d_x$ and $d_y$. It
can be shown that
\begin{equation}\label{Dek}
[D,\varepsilon_{\mk}]=\frac{i\hbar^2}{m}H_{so}.
\end{equation}
Using the unitary transformation Eq. (\ref{Unitary}) and Eq.
(\ref{Dek}), then Eq. (\ref{Hamiltonian}) becomes (up to the
second order of $\beta_{ij}$)
\begin{equation}\label{UHU-1}
\begin{split}
H'=&U^{\dag}HU\\
=&\varepsilon_{\mk}+\frac{im}{\hbar^2}\left(1-\frac{1}{2!}\right)[D,H_{so}]+V(x,y)+o(\beta^3_{ij}).
\end{split}
\end{equation}
It can be shown that
\begin{equation}\label{DHso}
[D,H_{so}]=i\sum_{ij}\beta^2_{ij}+\frac{2i}{\hbar}\left(\detbeta\right)\sigma_zL_z,\\
\end{equation}
where $L_z=\hbar(xk_y-yk_x)$ is the orbital angular momentum.
Substitute Eq. (\ref{DHso}) into Eq. (\ref{UHU-1}), we obtain
\begin{equation}\label{UHU}
H'=\varepsilon_{\mk}-\frac{m}{2\hbar^2}\sum_{ij}\beta^2_{ij}+h_D+V(x,y)+o(\beta^3_{ij}),
\end{equation}
where
\begin{equation}\label{UHU-D}
h_D=-\frac{m}{\hbar^3}\left(\detbeta\right)\sigma_zL_z.
\end{equation}
Equation (\ref{UHU-D}) shows that the coupling between orbital
motion $L_z$ and spin $z$ component $\sigma_z$ is proportional to
$\detbeta$. Therefore, Eq. (\ref{detbeta}) together with Eqs.
(\ref{SHC-1}) and (\ref{UHU-D}) exhibits a discriminant for a
non-vanishing spin-Hall conductivity:
\begin{equation}
\begin{split}
&\detbeta=0\rightarrow\sigma^z_{xy}=0,\\
&\detbeta\neq0\rightarrow\sigma^z_{xy}\neq0.
\end{split}
\end{equation}
In the Rashba-Dresselhaus system
($\widetilde{\beta}_R+\widetilde{\beta}_S$), we have
$\beta_{xx}=\beta$, $\beta_{xy}=\alpha$, $\beta_{yx}=-\alpha$,
$\beta_{yy}=-\beta$, and $\detbeta=\alpha^2-\beta^2$.  It has been
shown that the the vanishing spin-Hall conductivity in the
Rashba-Dresselhaus system results from the fact that the orbital
motion is decoupled from the spin $z$-component when
$\alpha^2=\beta^2$.~\cite{Manuel11}

On the other hand, band degeneracy occurs when
$k_F^+(\phi^*)=k_F^-(\phi^*)$, namely, the inner band and outer
band overlap for some vale $\phi^*$. The solution $\phi^*$ is
given by
\begin{equation}\label{phistar}
\tan\phi^*=\frac{-(\beta_{xx}\beta_{xy}+\beta_{yx}\beta_{yy})\pm\sqrt{-(\detbeta)^2}}{\beta_{xy}^2+\beta_{yy}^2}.
\end{equation}
If $\detbeta\neq0$, the term $\sqrt{-(\detbeta)^2}$ is a complex
number and the angle $\phi^*$ does not exist. The angle $\phi^*$
exists only when $\detbeta=0$. The degeneracy could be open upon
tuning the spin-orbit interactions such that $\detbeta\neq0$.
Therefore, the decoupling of the spin $s_z$ and orbital motion
$L_z$ always accompanies the band-overlapping phenomenon. The
decoupling of spin and orbital motion results in a sign change of
the ISHC and the band-overlapping phenomenon.

\section{Asymmetric and the upper bound value of ISHC}\label{sec:MVSHC}

In order to evaluate the integral in Eq. (\ref{SHC-1}), we
transform the integral to the contour integral in a complex plane.
If $z$ is defined as $z=e^{i\phi}$, the integral becomes a line
integral along a closed loop with unit radius. The function
$\Gamma(\phi)$ can be rewritten as
\begin{equation}
\Gamma(\phi)^2\rightarrow\Gamma(z)^2=\frac{1}{4z^2}\left(\lambda_1z^2+\lambda_2\right)\left(\lambda_2^*z^2+\lambda_1^*\right),
\end{equation}
where $"*"$ symbolizes the complex conjugate and

\begin{equation}\label{lambdas}
\begin{split}
&\lambda_1=(\beta_{xx}+\beta_{yy})+i(\beta_{yx}-\beta_{xy})\\
&\lambda_2=(\beta_{xx}-\beta_{yy})+i(\beta_{yx}+\beta_{xy}).
\end{split}
\end{equation}
The integral in Eq. (\ref{SHC-1}) can be evaluated by calculating
the residue inside the unit circle $|z|=1$. The conditions for the
poles appearing in the unit circle indicate the boundary of change
of ISHC in sign. By using the standard residue theorem
\cite{arf95}, the result is derived as
\begin{equation}\label{SHC-2}
\int_0^{2\pi}d\phi\frac{\cos^2\phi}{\Gamma(\phi)^2}=\frac{4\pi}{|\lambda_>|^2-|\lambda_<|^2}\left[1-\mathrm{Re}\left(\frac{\lambda_<}{\lambda_>}\right)\right],
\end{equation}
where $\mathrm{Re}(\cdots)$ represents the real part of a complex
number. $\lambda_>$ ($\lambda_<$) is taken from the relative
maximum (minimum) value of ($|\lambda_1|$, $|\lambda_2|$). That
is, if $|\lambda_1|>|\lambda_2|$ then $\lambda_>=\lambda_1$ and
$\lambda_<=\lambda_2$, and vice versa. Equation (\ref{SHC-2}) can
be further simplified. Using Eq. (\ref{lambdas}), we find that
\begin{equation}\label{SHC-3}
|\lambda_1|^2-|\lambda_2|^2=4\detbeta.
\end{equation}
If $\detbeta\neq0$ in Eq. (\ref{SHC-3}), it cancels $\detbeta$
appearing in Eq. (\ref{SHC-1}). Combining Eq. (\ref{SHC-1})
together with (\ref{SHC-2}) and (\ref{SHC-3}), we have
\begin{equation}\label{SHC-result}
\sigma^z_{xy}=\left\{\begin{array}{cc}
\displaystyle \mathrm{sgn}(\detbeta)\frac{e}{8\pi}\left[1-\mathrm{Re}\left(\frac{\lambda_<}{\lambda_>}\right)\right]&;\detbeta\neq0\\
\\
0&;\detbeta=0\\
\end{array}\right.,
\end{equation}
where $\mathrm{sgn}(\detbeta)=\detbeta/|\detbeta|$ is the sign
function. We have $\mathrm{sgn}(\detbeta>0)=+1$ and
$\mathrm{sgn}(\detbeta<0)=-1$. The real part of
$\lambda_</\lambda_>$ in Eq. (\ref{SHC-result}) can be written in
terms of coefficients of spin-orbit matrix,
\begin{equation}\label{Relambda}
\mathrm{Re}\left(\frac{\lambda_<}{\lambda_>}\right)=\frac{(\beta_{xx}^2-\beta_{yy}^2)+(\beta_{yx}^2-\beta_{xy}^2)}{\sum_{ij}\beta_{ij}^2+2|\detbeta|}.
\end{equation}
Note that in Eq. (\ref{Relambda}), there is an absolute value of
$\detbeta$. The ISHC generally depends on the spin-orbit
interaction [Eq. (\ref{Relambda})], and it is not necessarily a
universal constant. We note that the denominator of Eq.
(\ref{Relambda}) is always positive. Nevertheless, the numerator
of Eq. (\ref{Relambda}) can be either positive or negative. For
convenience in the following discussion, we define the asymmetric
response function $\Deltabeta$ as
\begin{equation}\label{Deltabeta}
\begin{split}
\Deltabeta&\equiv(\beta_{xx}^2-\beta_{yy}^2)+(\beta_{yx}^2-\beta_{xy}^2)\\
&=\Gamma(0)^2-\Gamma(\pi/2)^2.
\end{split}
\end{equation}
We find that the asymmetric response function involves two
quantities: $2m\Gamma(0)/\hbar^2$ is the band-splitting along the
direction of the spin-Hall response and $2m\Gamma(\pi/2)/\hbar^2$
is the band-splitting along the direction of the applied electric
field [see Eq. (\ref{diff-FM})]. The asymmetric response function
is the difference of two specific band-splittings.

For $\Deltabeta\geq0$, the ISHC is less than $e/8\pi$. Therefore,
the spin-Hall conductivity has an upper bound in magnitude,
\begin{equation}\label{D1}
|\sigma^z_{xy}|\leq\frac{e}{8\pi},~\Deltabeta\geq0.
\end{equation}
The equality in Eq. (\ref{D1}) is valid only when
$(\beta_{xx}^2-\beta_{yy}^2)+(\beta_{yx}^2-\beta_{xy}^2)=0$ in
coordinate system $(k_x,k_y)$. If $k_x$ axis is along [100]
direction and $k_y$ axis is along [010] direction, then some
spin-orbit coupled systems would satisfy this condition, for
example, the pure Rashba, the pure Dresselhaus, and the
Rashba-Dresselhaus systems. This result is in agreement with the
previous theoretical results~\cite{RD}.

\begin{figure}
\begin{center}
\includegraphics[width=6cm,height=4.5cm]{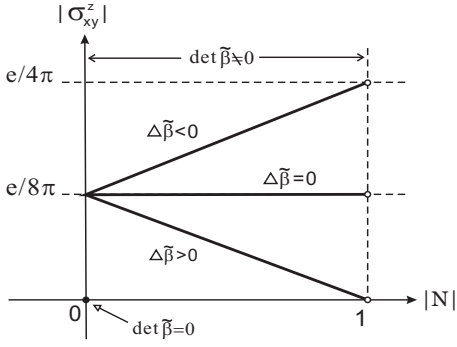}
\end{center}
\caption{Variation in the spin-Hall conductivity $|\sigma^z_{xy}|$
with $|N|$ for three conditions $\Deltabeta>0$, $\Deltabeta=0$,
and $\Deltabeta<0$, where $Re(\lambda_</\lambda_>)$ is defined as
$N$ and $|N|<1$. }\label{fig1}
\end{figure}

Interestingly, we find that if $\Deltabeta<0$, then
$\mathrm{Re}\left(\lambda_</\lambda_>\right)$ is negative and the
spin-Hall conductivity satisfies
\begin{equation}
\frac{e}{8\pi}<|\sigma^{z}_{xy}|<\frac{e}{4\pi},~\Deltabeta<0.
\end{equation}
The ISHC still has an upper bounded value $e/4\pi$; however, it
can exceed the value $e/8\pi$. The three conditions are summarized
in \figref{fig1}, where we define
$N=\mathrm{Re}(\lambda_</\lambda_>)$ and it has been shown that
$|N|<1$. The spin-Hall conductivity
$|\sigma^z_{xy}|=(e/8\pi)(1-|N|)$ for $\Deltabeta>0$,
$|\sigma^z_{xy}|=e/8\pi$ for $\Deltabeta=0$ and
$|\sigma^z_{xy}|=(e/8\pi)(1+|N|)$ for $\Deltabeta<0$.

When an electric field is applied in the $k_x$ direction, the
spin-Hall response in the $k_y$ direction is given by
\begin{equation}\label{SHCyx}
\sigma^z_{yx}=-\frac{e}{8\pi^2}\left(\detbeta\right)\int_0^{2\pi}d\phi\frac{\sin^2\phi}{\Gamma(\phi)^2}.
\end{equation}
The integration in Eq. (\ref{SHCyx}) can also be calculated
analytically as follows:
\begin{equation}\label{SHCyx_2}
\sigma^z_{yx}=\left\{\begin{array}{cc}
\displaystyle -\mathrm{sgn}(\detbeta)\frac{e}{8\pi}\left[1+\mathrm{Re}\left(\frac{\lambda_<}{\lambda_>}\right)\right]&;\detbeta\neq0\\
\\
0&;\detbeta=0\\
\end{array}\right.
\end{equation}
Unlike the $\sigma^z_{xy}$, we have
\begin{equation}
\begin{split}
&|\sigma^z_{yx}|\leq\frac{e}{8\pi},~\Deltabeta\leq0.\\
&\frac{e}{8\pi}<|\sigma^{z}_{xy}|<\frac{e}{4\pi},~\Deltabeta>0.
\end{split}
\end{equation}
We find that when $|\sigma_{xy}^z|$ is larger than $e/8\pi$,
$|\sigma_{yx}^z|$ is less than $e/8\pi$ and vice versa. In
comparison with $\sigma^z_{xy}$ [Eq. (\ref{SHC-result})], we find
that $\sigma^z_{xy}$ is in general not equal to $-\sigma^z_{yx}$
in the k-linear system.

The symmetrical result ($\sigma^z_{xy}=-\sigma^z_{yx}$) is
obtained because the electric field is applied in a direction such
that the asymmetric response function vanishes. Both the pure
Rashba and the pure Dresselhaus systems exhibit circular energy
dispersion, and the asymmetric response function always vanishes
regardless of the direction of the applied electric field. In the
Rashba-Dresselhaus system, if the electric field is applied along
the direction of $[010]$ and the spin-Hall response occurs along
$[100]$, the band splitting along the direction of the applied
electric field is the same as that along the spin-Hall response
direction, and thus, the asymmetric response function vanishes.
However, a small change in the direction of the applied electric
field would result in a non-vanishing asymmetric response function
in the Rashba-Dresselhaus system. The influence of the applied
electric field on the asymmetric response function is discussed in
the next section.

\section{Maximum value of ISHC and asymmetric response function}\label{sec:A-ISHC}

The direction of the applied electric field plays an important
role in determining whether the system has an non-vanishing
asymmetric response function. We study the asymmetric response
function by rotating the coordinate system from $(k_x,k_y)$ to
$(k_x',k_y')$. Consider counterclockwise rotation of the system
along the along $z$ axis by an angle $\Theta$; the relationship
between $(k_x,k_y)$ and $(k_x',k_y')$ is given by
$k_x=k_x'\cos\Theta-k_y'\sin\Theta$ and
$k_y=k_x'\sin\Theta+k_y'\cos\Theta$. The term $\beta_{i'j'}$
represents the matrix element of the spin-orbit matrix in the new
coordinate, and they are given by
$\beta_{x'x'}=\beta_{xx}\cos\Theta+\beta_{xy}\sin\Theta$,
$\beta_{x'y'}=-\beta_{xx}\sin\Theta+\beta_{xy}\cos\Theta$,
$\beta_{y'x'}=\beta_{yx}\cos\Theta+\beta_{yy}\sin\Theta$, and
$\beta_{y'y'}=-\beta_{yx}\sin\Theta+\beta_{yy}\cos\Theta$. It can
be shown that the value of $\detbeta$ is independent of the choice
of coordinates, i.e.,
$\detbeta=(\beta_{xx}\beta_{yy}-\beta_{yx}\beta_{xy})=(\beta_{x'x'}\beta_{y'y'}-\beta_{y'x'}\beta_{x'y'})$.
Interestingly, it can also be shown that
$\sum_{ij}\beta^2_{i'j'}=\sum_{ij}\beta^2_{ij}$, namely,
$\sum_{ij}\beta^2_{ij}$ is also independent of the choice of
coordinates.

In the coordinate system $(k_x',k_y')$, the term $\sigma^z_{x'y'}$
indicates that the electric field is applied along the $k_y'$
direction ($\phi'=\pi/2$) and the corresponding spin-Hall response
$J^z_{x'}$ is along the $k_x'$ direction ($\phi'=0$), where the
angle $\phi'$ is measured from the positive axis of $k_x'$. On the
other hand, the term $\sigma^z_{y'x'}$ indicates that the electric
field is applied along the $k_x'$ direction ($\phi'=0$) and the
corresponding spin-Hall response $J^z_{y'}$ is obtained along the
$k_y'$ direction ($\phi'=\pi/2$). Therefore, in the new coordinate
system, Eqs. (\ref{SHC-result}) and (\ref{SHCyx_2}) are still
valid and can be written as
\begin{equation}\label{SHC-newc}
\begin{split}
&\sigma^z_{x'y'}=\mathrm{sgn}(\detbeta)\frac{e}{8\pi}\left[1-\mathrm{Re}\left(\frac{\lambda'_<}{\lambda'_>}\right)\right],\\
&\sigma^z_{y'x'}=-\mathrm{sgn}(\detbeta)\frac{e}{8\pi}\left[1+\mathrm{Re}\left(\frac{\lambda'_<}{\lambda'_>}\right)\right],
\end{split}
\end{equation}
where
\begin{equation}\label{SHC-newc2}
\mathrm{Re}\left(\frac{\lambda'_<}{\lambda'_>}\right)=\frac{\Gamma'(\phi'=0,\Theta)^2-\Gamma'(\phi'=\frac{\pi}{2},\Theta)^2}{\sum_{ij}\beta^{2}_{ij}+2|\detbeta|}.
\end{equation}
The energy dispersion in the new coordinate system is
$\Gamma'(\phi',\Theta)^2=(\beta_{x'x'}^2+\beta_{y'x'}^2)\cos^2\phi'+(\beta_{x'y'}^2+\beta^2_{y'y'})\sin^2\phi'+(\beta_{x'x'}\beta_{x'y'}+\beta_{y'x'}\beta_{y'y'})\sin(2\phi')$.
Equation (\ref{SHC-newc2}) indicates that the variation in the
ISHC is altered only by the difference in two band-splittings: the
band-splitting along the applied electric field direction and the
band-splitting along the spin-Hall response direction.

It can be shown that in the new coordinate system,
$\Gamma'(0,\Theta)$ and $\Gamma'(\pi/2,\Theta)$ can be written as
\begin{equation}\label{GammaArb}
\begin{split}
\Gamma'(0,\Theta)^2=&\Gamma(0)^2\cos^2\Theta+\Gamma(\pi/2)^2\sin^2\Theta\\
&+(\beta_{xx}\beta_{xy}+\beta_{yx}\beta_{yy})\sin(2\Theta),\\
\Gamma'(\pi/2,\Theta)^2=&\Gamma(0)^2\sin^2\Theta+\Gamma(\pi/2)^2\cos^2\Theta\\
&-(\beta_{xx}\beta_{xy}+\beta_{yx}\beta_{yy})\sin(2\Theta),
\end{split}
\end{equation}
where $\Gamma(0)^2=\beta_{xx}^2+\beta_{yx}^2$ and
$\Gamma(\pi/2)^2=\beta_{xy}^2+\beta_{yy}^2$. Therefore, in
general, when $\Theta\neq0$, $\Gamma'(0,\Theta)$ is not equal to
$\Gamma'(\pi/2,\Theta)$, even if $\Gamma(0)=\Gamma(\pi/2)$. A
small rotation of the direction of the applied electric field
would lead to a non-vanishing asymmetric response function.

According to Eqs. (\ref{SHC-newc}) and (\ref{SHC-newc2}), in order
to enhance $\sigma^z_{x'y'}$, i.e., $|\sigma^z_{x'y'}|>e/8\pi$,
the band-splitting must satisfy the condition
$\Gamma'(0,\Theta)<\Gamma'(\pi/2,\Theta)$. This means that the
electric field must be applied along the direction with the larger
band-splitting in comparison with that in the direction of the
spin-Hall response. On the other hand, the corresponding vaalue of
$\sigma^z_{y'x'}$ is less than $e/8\pi$. Conversely, if we want to
enhance $\sigma^z_{y'x'}$, i.e. $|\sigma^z_{y'x'}|>e/8\pi$, then
we must have $\Gamma'(0,\Theta)>\Gamma'(\pi/2,\Theta)$. This means
that the electric field must be applied along the direction with
larger band-splitting in comparison with that in the direction of
the spin-Hall response.

Therefore, we conclude that in order to obtain the ISHC $\sigma_s$
with $\sigma_s>e/8\pi$ ($\sigma_s$ can be $|\sigma_{x'y'}^z|$ or
$|\sigma_{y'x'}^z|$), the band splitting along the direction of
the applied electric field must be larger than that along the
direction of the spin-Hall response.

As indicated in \secref{sec:A-ISHC}, when $\sigma_s>e/8\pi$,
$\sigma_s$ still has an upper bound value of $e/4\pi$. This means
that we can further enhance $\sigma_s$ by finding the maximum
value of the asymmetric response function. In fact, it can be
shown that (see Appendix~\ref{Appendix:MNBS}) when all the
strengths of the spin-orbit interactions are fixed, the maximum
value of $|\Gamma'(0,\Theta)^2-\Gamma'(\pi/2,\Theta)^2|$ exists
for a specific direction $\Theta_M$ [see Eq. (\ref{ThetaM})]. This
also implies that the direction of the largest band-splitting is
always perpendicular to that of the smallest band-splitting on the
Fermi surface. Furthermore, the existence of the maximum value
$|\Gamma'(0,\Theta)^2-\Gamma'(\pi/2,\Theta)^2|$ on the Fermi
surface provides us a method to obtain the maximum ISHC with
respect to these fixed values of $\beta_{ij}$.

In the next section, we explain how the enhanced spin-Hall
response is the manifestation of Berry curvature at the nearly
degenerate area.

\section{Berry curvature and the nearly degenerate area}\label{sec:BerryC}
\begin{figure}[!htb]
\begin{center}
\includegraphics[width=8cm,height=3.5cm]{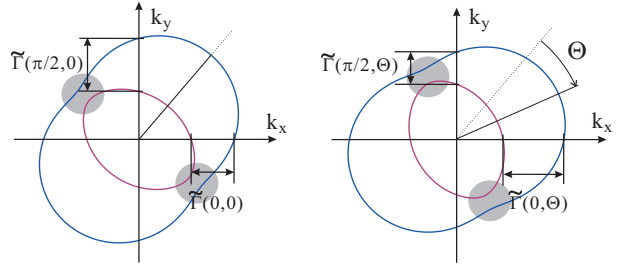}
\end{center}
\caption{(Color online) Rotation of energy dispersions described
by $\Theta$. The band splitting $(2m/\hbar^2)\Gamma(\phi,\Theta)$
is defined as
$(2m/\hbar^2)\Gamma(\phi,\Theta):=\widetilde{\Gamma}(\phi,\Theta)$.
The circular area (gray) represents the nearly degenerate area.
 }\label{figRotation}
\end{figure}

In this section, we analyze the Berry curvature in a system with
fixed spin-orbit interactions, and for convenience, the direction
of the electric field is fixed while we rotate the system (see
\figref{figRotation}).

The spin-Hall conductivity $\sigma^z_{xy}$ [Eq. (\ref{SHC})] can
be written in terms of the Berry curvature as
\begin{equation}\label{SHC_Berry}
\sigma_{xy}^z=-\frac{e}{\hbar}\frac{1}{V}\sum_{\mk}\sum_{n=\pm}f_{\mk
n}\Omega^{(n)}_{xy}(\mk),
\end{equation}
and it can be shown that
\begin{equation}\label{BerryC}
\begin{split}
&\Omega^{(+)}_{xy}(\mk)=+\frac{\hbar^3}{4m}\detbeta\frac{\cos^2\phi}{k\Gamma(\phi,\Theta)^3},\\
&\Omega^{(-)}_{xy}(\mk)=-\frac{\hbar^3}{4m}\detbeta\frac{\cos^2\phi}{k\Gamma(\phi,\Theta)^3}.
\end{split}
\end{equation}
The energy dispersion $\Gamma(\phi,\Theta)$ is given by
$\Gamma(\phi,\Theta)^2=A\cos^2\phi+B\sin^2\phi+C\sin(2\phi)$,
where
$A=(\beta_{xx}^2+\beta_{yx}^2)\cos^2\Theta+(\beta_{xy}^2+\beta_{yy}^2)\sin^2\Theta+(\beta_{xx}\beta_{xy}+\beta_{yx}\beta_{yy})\sin(2\Theta)$,
$B=(\beta_{xx}^2+\beta_{yx}^2)\sin^2\Theta+(\beta_{xy}^2+\beta_{yy}^2)\cos^2\Theta-(\beta_{xx}\beta_{xy}+\beta_{yx}\beta_{yy})\sin(2\Theta)$,
and
$C=(\beta_{xy}^2-\beta_{xx}^2+\beta_{yy}^2-\beta_{yx}^2)\sin(2\Theta)/2+(\beta_{xx}\beta_{xy}+\beta_{yx}\beta_{yy})\cos(2\Theta)$.

We note that the Berry curvature of outer band ($n=+1$) is
opposite to that of the inner band ($n=-1$) in sign. Therefore,
when both bands are occupied ($\mu>0$), the only contribution to
the spin-Hall conductivity is the Berry curvature of the outer
band. Namely, at a fixed value of $\phi$, the Berry curvature of
the inner band cancels that of the outer band at every $k$ point
with $k<k_F^-$. Equation (\ref{SHC_Berry}) becomes
\begin{equation}\label{SigmaOmegaxy}
\sigma_{xy}^z=\frac{e}{\hbar}\int_{k_F^-}^{k_F^+}dS_k\Omega^{(+)}_{xy}(\mk).
\end{equation}
On the other hand, for $\sigma_{yx}^z$, we have
\begin{equation}\label{SigmaOmegayx}
\begin{split}
\sigma_{yx}^z=-\frac{e}{\hbar}\int_{k_F^-}^{k_F^+}dS_k\Omega^{(+)}_{yx}(\mk),
\end{split}
\end{equation}
where
\begin{equation}
\Omega^{(+)}_{yx}(\mk)=+\frac{\hbar^3}{4m}\detbeta\frac{\sin^2\phi}{k\Gamma(\phi,\Theta)^3}.
\end{equation}
We plot the variation in the Berry curvatures
$\Omegaxy:=\Omega_{xy}^{(+)}/(\hbar^3/4m)$ and
$\Omegayx:=\Omega_{yx}^{(+)}/(\hbar^3/4m)$ along the path
$k_F^+(\phi)=m\Gamma(\phi,\Theta)/\hbar^2+\sqrt{(m/\hbar^2)^2\Gamma(\phi,\Theta)^2+(2m\mu/\hbar^2)^2}$.
The direction of the electric field is fixed along $k_y$ for
obtaining $\Omegaxy$ or along $k_x$ for obtaining $\Omegayx$. The
peak value of $\Omegaxy$ ($\Omegayx$) is denoted as $\MaxOmegaxy$
($\MaxOmegayx$). The peak value refers to the value of the Berry
curvature at the nearly degenerate area (see \figref{figBC}). The
ISHCs are $|\sigma^z_{xy}|=(e/8\pi)(1-N)$ and
$|\sigma^z_{yx}|=(e/8\pi)(1+N)$, where
$N=[\Gamma(0,\Theta)^2-\Gamma(\pi/2,\Theta)^2]/(\sum_{ij}\beta_{ij}^2+2|\detbeta|)$.

We select a system with a non-spherical energy dispersion (the
nearly degenerate area exists) and
$\Gamma(0,0)\neq\Gamma(\pi/2,0)$. We use the following
coefficients of the spin-orbit matrix:
$\beta_{xx}=6\times10^{-2}$~eV~nm,
$\beta_{xy}=4.14\times10^{-2}$~eV~nm,
$\beta_{yx}=4.85\times10^{-2}$~eV~nm,
$\beta_{yy}=8.71\times10^{-2}$~eV~nm. The Fermi energy is 2.67
meV. The particle mass is 0.08 in units of the bare electron mass.
The energy dispersion is non-spherical, as shown in the left-hand
side of \figref{figBC}(a). The thick line represents the direction
with the largest band-splitting on the Fermi surface. By using Eq.
(\ref{ThetaM}), we have $\Theta_{M1}\sim0.288922\pi$ and
$\Theta_{M2}\sim0.788922\pi$. The asymmetric response function
vanishes at $\Theta=\Theta_0$, and it can be obtained by using
$\Gamma(0,\Theta_0)=\Gamma(\pi/2,\Theta_0)$. The result is
$\Theta_{01}\sim0.038922\pi$ and $\Theta_{02}\sim0.538921\pi$.

\begin{figure}[!htb]
\begin{center}
\includegraphics[width=8cm,height=11cm]{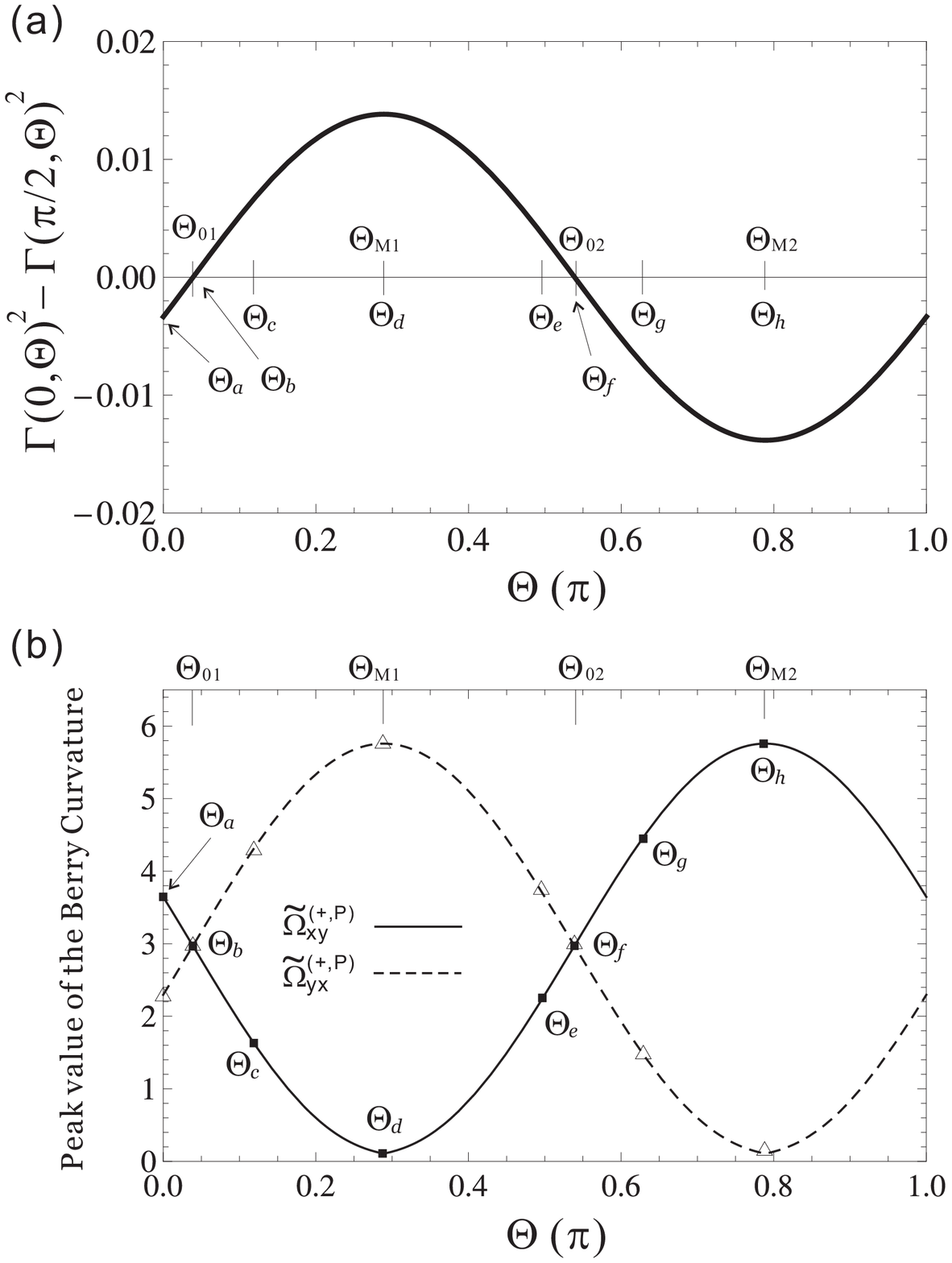}
\end{center}
\caption{ (a) The asymmetric response function
[$\Gamma(0,\Theta)^2-\Gamma(\pi/2,\Theta)^2$ (eV~cm)$^2$] vs
$\Theta$. (b) Variation of the peak values of the Berry curvatures
($\MaxOmegaxy$ and $\MaxOmegayx$) with $\Theta$. The Berry
curvatures have units of $(\hbar^3/4m)\times10^6$. The peak values
$\MaxOmegaxy$ ($\blacksquare$) and $\MaxOmegayx$ ($\vartriangle$)
correspond to \figref{figBC}(a) to (h).
 }\label{figPeakBC}
\end{figure}
The variations of asymmetric response function and the peak values
of the Berry curvatures with $\Theta$ are shown in
\figref{figPeakBC}. When the system is not rotated ($\Theta=0$),
the band splitting at $\phi=0$ is less than that at $\phi=\pi/2$
[\figref{figPeakBC}(a)], and we have
$|\sigma^z_{xy}|>|\sigma^z_{yx}|$. We find that $\MaxOmegaxy$ is
larger than $\MaxOmegayx$ as shown in \figref{figPeakBC}(b).

We now rotate the system with an angle $\Theta=\Theta_{01}$ such
that the band splitting at $\phi=0$ is equal to that at
$\phi=\pi/2$, resulting a vanishing asymmetric response function,
i.e., $\Gamma(0,\Theta_{01})=\Gamma(\pi/2,\Theta_{01})$
[\figref{figPeakBC}(a)]. This results in
$|\sigma^z_{xy}|=|\sigma_{yx}^z|=e/8\pi$. In
\figref{figPeakBC}(b), $\MaxOmegaxy$ is equal to $\MaxOmegayx$
when $\Theta=\Theta_{01}$. If we further rotate the system such
that the band splitting at $\phi=0$ is larger than that at
$\phi=\pi/2$ ($\Theta_{01}<\Theta<\Theta_{M1}$), we obtain
$|\sigma_{xy}^z|<|\sigma_{yx}^z|$. The corresponding $\MaxOmegayx$
is now larger than $\MaxOmegaxy$ as can be seen in
\figref{figPeakBC}(b). If we rotate the system by
$\Theta=\Theta_{M1}$ such that the largest band splitting is now
located along the $k_x$ direction, the magnitude of the asymmetric
response function in this case reaches a maximum value as shown in
\figref{figPeakBC}(a). We still have
$|\sigma_{xy}^z|<|\sigma_{yx}^z|$, but $|\sigma_{yx}^z|$ would be
very close to $e/4\pi$, and $|\sigma^z_{xy}|$ is less than
$e/8\pi$. We find that $\MaxOmegayx$ is not only considerably
larger than that of $\MaxOmegaxy$, but is also the maximum value
in comparison with the other peak values of the Berry curvatures.

When $\Theta=\Theta_{02}$, the asymmetric response function
vanishes. In this case, we obtain
$|\sigma_{xy}^z|=|\sigma_{yx}^z|=e/8\pi$ and
$\MaxOmegaxy=\MaxOmegayx$. When the maximum peak value of the
Berry curvature $\MaxOmegaxy$ ($\MaxOmegayx$) is obtained at
$\Theta=\Theta_{M2}$ ($\Theta=\Theta_{M1}$), the magnitude of the
asymmetric response function reaches a maximum value. We have
$|\sigma_{xy}^z|>|\sigma_{yx}^z|$, but $\sigma_{xy}^z$ would be
very close to $e/4\pi$, and $\sigma_{yx}$ is less than $e/8\pi$.

We select some specific angles [$\Theta_a\sim\Theta_h$ shown in
\figref{figPeakBC}(a) and (b)] and plot the corresponding Berry
curvatures along the path $k_F^+(\phi)$ and the orientations of
the energy dispersions (see \figref{figBC}).

As shown in \figref{figBC}, both the Berry curvatures ($\Omegaxy$
and $\Omegayx$) have significant values (peak values) in the
nearly degenerate area. When the system is rotated, both the
positions of nearly degenerate area and $\MaxOmegaxy$ and
$\MaxOmegayx$ change together. Furthermore, $\MaxOmegayx$
increases and $\MaxOmegaxy$ decreases as seen from
\figref{figBC}(a) to (d). On the other hand, in \figref{figBC}(e)
to (h), $\MaxOmegayx$ decreases and $\MaxOmegaxy$ increases [see
also \figref{figPeakBC}(b)].

\begin{widetext}

\begin{figure}[!htb]
\begin{center}
\includegraphics[width=16cm,height=14cm]{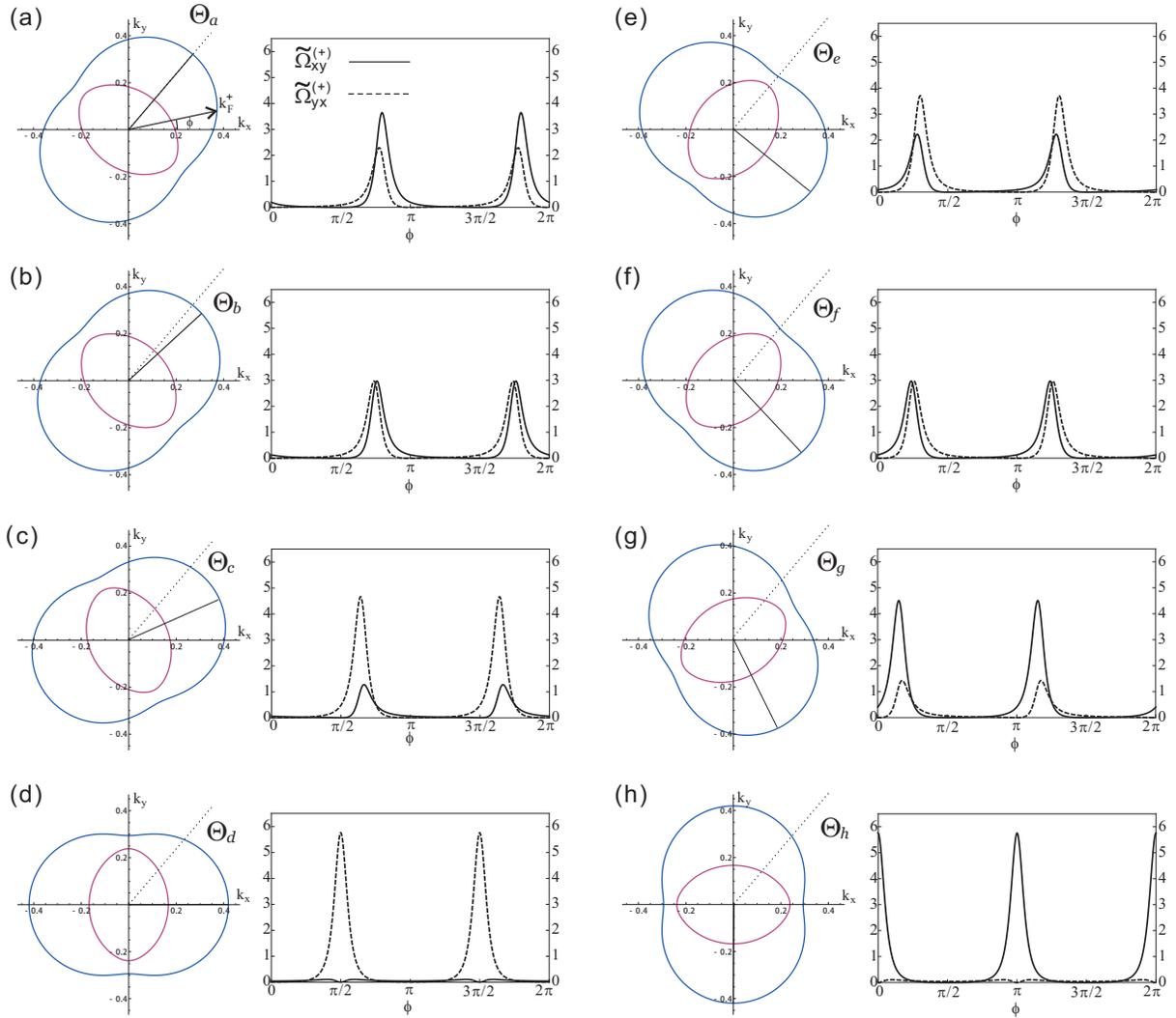}
\end{center}
\caption{(Color online) Rotation of energy dispersions described
by $\Theta$ and the corresponding variation in the Berry curvature
$\Omegaxy$ (thick line) and $\Omegayx$ (broken line) along the
path $k_F^+(\phi)$, $\phi:0\sim2\pi$. Both the Berry curvatures
($\Omegaxy$ and $\Omegayx$) have units of
$(\hbar^3/4m)\times10^6$. (a)$\Theta_a=0$,
(b)$\Theta_b=\Theta_{01}$, (c)$\Theta_{01}<\Theta_c<\Theta_{M1}$,
(d)$\Theta_d=\Theta_{M1}$, (e)$\Theta_{M1}<\Theta_e<\Theta_{02}$,
(f)$\Theta_f=\Theta_{02}$, (g)$\Theta_{02}<\Theta_g<\Theta_{M2}$,
(h)$\Theta_h=\Theta_{M2}$.}\label{figBC}
\end{figure}

\end{widetext}

\begin{figure}[!htb]
\begin{center}
\includegraphics[width=7cm,height=10cm]{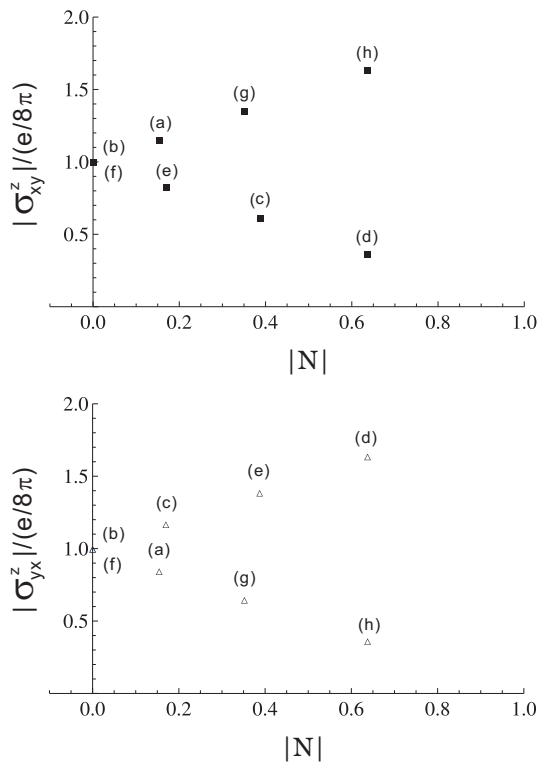}
\end{center}
\caption{The numerical values of ISHCs $\sigma^z_{xy}$
($\blacksquare$) and $\sigma_{yx}^z$ ($\vartriangle$)
corresponding to different rotation angles $\Theta$ corresponding
to \figref{figBC}(a) to (h) and $\Theta_a\sim\Theta_h$ in
\figref{figPeakBC}.}\label{figSHCvsN}
\end{figure}

The variation in the ISHCs ($|\sigma_{xy}^z|$ and
$|\sigma_{yx}^z|$) corresponding to the Berry curvature variations
in \figref{figBC}(a) to (h) is shown in \figref{figSHCvsN}, where
$|\sigma^z_{xy}|=(e/8\pi)(1-N)$, $|\sigma^z_{yx}|=(e/8\pi)(1+N)$
and
$N=[\Gamma(0,\Theta)^2-\Gamma(\pi/2,\Theta)^2]/(\sum_{ij}\beta_{ij}^2+2|\detbeta|)$.
The ISHC $|\sigma_{xy}^z|$ decreases in \figref{figSHCvsN}(a) to
(d) and then increases in \figref{figSHCvsN}(e) to (h). The ISHC
$|\sigma_{yx}^z|$ increases in \figref{figSHCvsN}(a) to (d) and
subsequently decreases in \figref{figSHCvsN}(e) to (h). In the
cases shown in \figref{figSHCvsN}(b) and (f) ,
$|\sigma_{xy}^z|=|\sigma_{yx}^z|$ and both the Berry curvatures
have the same peak values. The behavior of the Berry curvature at
the nearly degenerate area is in agreement with our conclusions in
\secref{sec:A-ISHC}.

In the case of spherical energy dispersion (there is no nearly
degenerate area), it can be shown that the Berry curvature
$\Omegaxy$ equivalent to $\Omegayx$ shifted by $\pi/2$. As in the
non-spherical case, the Berry curvature still exhibits two
significant responses along the directions of the spin-Hall
current, but the shape and peak value of the Berry curvature do
not change when we rotate the system. This means that
$|\sigma^z_{xy}|=|\sigma^z_{yx}|$ regardless of the orientation of
the system. The magnitude of the asymmetric response function
always vanishes in this case, and the ISHC is a universal constant
$e/8\pi$.

It must be emphasized that the angle $\Theta_M$ enables us to find
the maximum value of the asymmetric response function for some
fixed $\beta_{ij}$. If we have another set of values
$\beta'_{ij}$, the corresponding maximum value of the asymmetric
response function is in general different from that with
$\beta_{ij}$. The magnitude of the ISHC may further be enhanced by
tuning the spin-orbit interactions to change the maximum value of
the magnitude of the asymmetric response function, but it still
has an upper bound of $e/4\pi$.

The measurable responses caused by the spin-Hall effect are very
different from those in the present idealized system, which is
infinite in size and does not include impurity scattering.
Measurable quantities such as spin accumulation, however, depend
on boundary conditions. The conserved spin-current considered in
the present paper may correspond to smooth
boundaries.~\cite{Shi06} However, the presence of impurities can
drastically affect clean limit
results~\cite{TWChen09Dis,Kh06,Burkov04,Nomura05}. In Ref.
\cite{Sherman10}, it was shown that impurity scattering does not
suppress the spin-Hall conductivity in the spatially random Rahsba
spin-orbit coupled system. In particular, the SU(2) formulation on
extrinsic mechanism of spin Hall conductivity was recently
investigated in Ref. \cite{Rai12}. However, the effects of a
finite size and impurity scattering are beyond the scope of the
present paper. Hopefully, our interesting predictions of higher
intrinsic ISHC would stimulate measurements in 2D semiconductor
systems in the near future.

\section{Equilibrium spin current and spin-orbit matrix}\label{sec:ESC}

We now turn to the discussion on equilibrium spin current in this
generic k-linear spin-orbit coupled system. In
Ref.~\cite{Rashba03}, it was shown that even in thermodynamic
equilibrium, spin current for the Rashba-Dresselhaus system does
not vanish in the absence of external fields.  This phenomena has
arisen many discussions on the definition of spin
current.~\cite{Rashba04,Sun07,Shi06,Sab08,Dro11} The possibilities
to detect the equilibrium spin currents have been studied in
Refs.~\cite{Sun07} and~\cite{Sonin07}.

We calculate the equilibrium spin current by using conventional
definition of spin current. In the case of the positive chemical
potential ($\mu>0$), two branches are populated. In the absence of
external fields, the equilibrium spin-current is the sum of the
in-plane spin currents of the two branches
\begin{equation}\label{ESC}
\langle J_i^{\sigma_j}\rangle=\frac{1}{V}\sum_{n\mk}f_{\mk
n}\langle n\mk|\frac{1}{2}\left\{\frac{\partial H}{\partial \hbar
k_i},\frac{\hbar}{2}\sigma_j\right\}|n\mk\rangle,
\end{equation}
where $i,j=x,y$ and $|n\mk\rangle$ is the eigenstate of
Hamiltonian Eq. (\ref{Hamiltonian}). From Eq. (\ref{diff-FM}) and
$k_{F}^+k_{F}^-=2m\mu/\hbar^2$, a straightforward calculation
yields
\begin{equation}\label{ESCR}
\left(\begin{array}{cc}
\langle J_x^{\sigma_x}\rangle&\langle J_x^{\sigma_y}\rangle\\
\langle J_y^{\sigma_x}\rangle&\langle J_y^{\sigma_y}\rangle
\end{array}\right)=N_F\left(\begin{array}{cc}
\beta_{yy}&-\beta_{xy}\\
-\beta_{yx}&\beta_{xx}
\end{array}\right)\detbeta,
\end{equation}
where $N_F=m^2/6\pi\hbar^4$. For specific systems, the result is
in agreement with the previous results. In the pure Rashba system,
where $\beta_{xy}=-\beta_{yx}=\alpha$ and
$\beta_{xx}=\beta_{yy}=0$, we have $\langle
J_y^{\sigma_x}\rangle=-\langle J_x^{\sigma_y}\rangle$ and $\langle
J_x^{\sigma_x}\rangle=\langle J_y^{\sigma_y}\rangle=0$. In the
pure Dresselhaus system, where $\beta_{xx}=-\beta_{yy}=\beta$ and
$\beta_{xy}=\beta_{yx}=0$, we have $\langle
J_x^{\sigma_x}\rangle=-\langle J_y^{\sigma_y}\rangle$ and $\langle
J_x^{\sigma_y}\rangle=\langle J_y^{\sigma_x}\rangle=0$. It is
interesting to note that the equilibrium spin current $\langle
J_i^{\sigma_j}\rangle$ is related to the inverse of spin-orbit
matrix $\widetilde{\beta}^{-1}$ via $\langle
J_i^{\sigma_j}\rangle=N_F(\widetilde{\beta}^{-1})_{ij}(\detbeta)^2$.

We find that $\detbeta$ also appears in the expression of
equilibrium spin current Eq. (\ref{ESCR}). However,
$J_i^{\sigma_j}$ occurs in the third order of $\beta_{ij}$. In
this sense, Eq. (\ref{UHU}) fails to explain the physical meaning
of $\detbeta$ in this case (see Appendix~\ref{Appendix:FW}).
Recently, the equilibrium spin current in k-linear spin-orbit
coupled systems is found to be link to the non-Abelian SU(2) gauge
theory, where the Pauli spin matrix serves as a color index in the
gauge field.~\cite{Tokatly08} The resulting color current
satisfies covariant conservation. The equilibrium spin current
obtained from the covariant conserved color current in the
Rashba-Dresselhaus systems is in agreement with
Ref.~\cite{Rashba03}. In the following, we apply this formalism to
the generic k-linear systems.

The Hamiltonian Eq. (\ref{Hamiltonian}) can be written in terms of
SU(2) gauge field $\hbar^2(\mk+\mathcal{A})^2/2m$, where
$\mathcal{A}=\mathbf{A}_i\hat{e}_i$  and
$\mathbf{A}_i=A_i^{a}\sigma_a$. We have
\begin{equation}\label{gf}
\begin{split}
&A_{x}^{x}=\frac{m}{\hbar^2}\beta_{xx},~A_{y}^{x}=\frac{m}{\hbar^2}\beta_{xy}\\
&A_{x}^{y}=\frac{m}{\hbar^2}\beta_{yx},~A_{y}^{y}=\frac{m}{\hbar^2}\beta_{yy}.
\end{split}
\end{equation}
The equilibrium spin current denoted as $J_i^a$~\cite{Tokatly08}
is proportional to $-i\epsilon_{abc}A^b_{i}F^c_{ij}$, where
$F^a_{ij}\sigma_a=-i[\mathbf{A}_i,\mathbf{A}_j]$ is the field
strength.

The physical meaning of $\detbeta$ in equilibrium spin current is
now clear. The field strength in the SU(2) non-Abelian gauge field
is given by~\cite{Tokatly08}
\begin{equation}
\begin{split}
F^a_{xy}\sigma_a&=-i[\mathbf{A}_x,\mathbf{A}_y]\\
&=-i[A_x^x\sigma_x+A_x^y\sigma_y,A_y^x\sigma_x+A_y^y\sigma_y]\\
&=-i\left(A_x^xA_y^y[\sigma_x,\sigma_y]+A_x^yA_y^x[\sigma_y,\sigma_x]\right)\\
&=2\sigma_z\frac{m^2}{\hbar^4}\left(\detbeta\right),
\end{split}
\end{equation}
where Eq. (\ref{gf}) is used. We have $F^x_{xy}=F^y_{xy}=0$ and
$F^z_{xy}=\frac{m^2}{\hbar^4}\left(\detbeta\right)$. That is,
$\detbeta$ plays the role of color magnetic field strength
$F^z_{xy}$.

\section{conclusions}\label{sec:con}
In conclusion, we have shown that in 2D and k-linear spin-orbit
coupled systems, the properties of the intrinsic spin-Hall
conductivity are governed by two quantities: the effective
coupling of spin and orbital motion (reflected by $\detbeta$) and
the asymmetric response function ($\Deltabeta$). The effective
coupling of spin and orbital motion is a discriminant for
determining whether or not the spin-Hall conductivity vanishes.
The decoupling of spin and orbital motion associated with
band-overlapping phenomenon explains the physical origin of the
sign change of the intrinsic spin-Hall conductivity.

Furthermore, the dependence of spin-orbit interaction on the
spin-Hall effect and the resulting unsymmetrical properties are
related to the asymmetric response function, which is determined
by the difference in band-splitting along two directions: those of
the applied electric field the and spin-Hall current. We varied
the orientation of the system and studied the variation in the
Berry curvature and the corresponding spin-Hall response. We found
that maximum intrinsic spin-Hall conductivity occurs along the
direction of the nearly degenerate area, which also leads to the
maximization of the Berry curvature and the magnitude of the
asymmetric response function. The position of the nearly
degenerate area can be determined analytically. We also showed
that the intrinsic spin-Hall conductivity has an upper bound value
of $e/4\pi$.

In addition, we showed that the equilibrium spin current is
proportional to $(\widetilde{\beta}^{-1})_{ij}(\detbeta)^2$, and
$\detbeta$ determines the field strength of the SU(2) non-Abelian
gauge field in equilibrium spin current.






\section*{ACKNOWLEDGMENTS}
We thank the National Science Council of Taiwan for the support
under Contract No. NSC 101-2112-M-110-013-MY3.

\appendix
\section{Foldy-Wouthuysen transformation}\label{Appendix:FW}
A unitary transformation can be generally written as $U=e^{-iS}$,
where the hermitian matrix $S$ can be expend in order of
$\beta_{ij}$, i.e. $S=S^{(1)}+S^{(2)}+S^{(3)}+\cdots$. That is,
$S^{(1)}$ represents the term proportional to the order of
$\beta_{ij}$, $S^{(2)}$ the order of $\beta^2_{ij}$ and so on.
Follow the approach of Foldy-Wouthuysen transformation, the
Hamiltonian $H=\varepsilon_{\mk}+H_{so}+V(x,y)$ under the unitary
transformation $U=e^{-iS}$ is given by
\begin{equation}
\begin{split}
H'&=e^{iS}He^{-iS}\\
&=\varepsilon_{\mk}+H^{(1)}+H^{(2)}+H^{(3)}+V(x,y)+o(\beta^4_{ij}),
\end{split}
\end{equation}
where
\begin{equation}\label{H12}
\begin{split}
&H^{(1)}=H_{so}+[iS^{(1)},\varepsilon_{\mk}]\\
&H^{(2)}=[iS^{(2)},\varepsilon_{\mk}]+[iS^{(1)},H_{so}]+\frac{1}{2!}[iS^{(1)},[iS^{(1)},\epsilon_{\mk}]]\\
\end{split}
\end{equation}
and
\begin{equation}\label{H3}
\begin{split}
H^{(3)}=&[iS^{(3)},\epsilon_{\mk}]+[iS^{(2)},H_{so}]+\frac{1}{2!}[iS^{(1)},[iS^{(2)},\varepsilon_{\mk}]]\\
&+\frac{1}{2!}[iS^{(2)},[iS^{(1)},\varepsilon_{\mk}]]+\frac{1}{2!}[iS^{(1)},[iS^{(1)},H_{so}]]\\
&+\frac{1}{3!}[iS^{(1)},[iS^{(1)},[iS^{(1)},\varepsilon_{\mk}]]].
\end{split}
\end{equation}
Because $H_{so}$ is an odd matrix, we have to find a matrix
$S^{(1)}$ to cancel this term. Namely, we require $H^{(1)}=0$ and
\begin{equation}
H_{so}+[iS^{(1)},\varepsilon_{\mk}]=0.
\end{equation}
On the other hand, we note that $H_{so}$ is made up of the linear
momentum $k_i$, i.e.,
$H_{so}=\sigma_x(\beta_{xx}k_x+\beta_{xy}k_y)+\sigma_y(\beta_{yx}k_x+\beta_{yy}k_y)$
and $\varepsilon_{\mk}$ is proportional to $k^2$, $S^{(1)}$ is
obtained by the replacements $k_x\rightarrow x$ and
$k_y\rightarrow y$ in $H_{so}$. Take into account the constant
$\frac{\hbar^2}{m}$, we have
\begin{equation}\label{S1}
iS^{(1)}=\frac{im}{\hbar^2}\left\{\sigma_x(\beta_{xx}x+\beta_{xy}y)+\sigma_y(\beta_{yx}x+\beta_{yy}y)\right\}.
\end{equation}
Substitute Eq. (\ref{S1}) into $H^{(2)}$, after a straightforward
calculation, we find that the last two terms
$[iS^{(1)},H_{so}]+\frac{1}{2!}[iS^{(1)},[iS^{(1)},\epsilon_{\mk}]]$
gives a diagonalized form
$i\sum_{ij}\beta^2_{ij}+\frac{2i}{\hbar}\left(\detbeta\right)\sigma_zL_z$.
This means that $H^{(2)}$ is already diagonalized, and thus, we
can choose
\begin{equation}\label{S2}
iS^{(2)}=0.
\end{equation}
Substitute Eqs. (\ref{S1}) and (\ref{S2}) into $H^{(3)}$ [Eq.
(\ref{H3})], we obtain
\begin{equation}
\begin{split}
H^{(3)}=&[iS^{(3)},\epsilon_{\mk}]\\
&-\frac{2m^2}{3\hbar^4}\left(\detbeta\right)[\{x,L_z\}(\sigma_y\beta_{xx}-\sigma_x\beta_{yx})\\
&+\{y,L_z\}(\sigma_y\beta_{xy}-\sigma_x\beta_{yy})].
\end{split}
\end{equation}
We find that the term in $[\cdots]$ is composed of odd matrices.
Therefore, we must require $H^{(3)}=0$. In this sense, the next
diagonalized part is order of $\beta^4_{ij}$.

\section{Maximum and minimum band-splitting}\label{Appendix:MNBS}

As shown in \secref{sec:MVSHC} and \secref{sec:A-ISHC}, the value
$\Gamma'(0,\Theta)^2-\Gamma'(\pi/2,\Theta)^2$ determines whether
the ISHC is larger than $e/8\pi$ or equal to $e/8\pi$.
Furthermore, if the magnitude
$|\Gamma'(0,\Theta)^2-\Gamma'(\pi/2,\Theta)^2|$ increases upon
varying $\Theta$, the ISHC would approach a maximum value with
respect to the fixed value of $\beta_{ij}$. We will show that the
maximum value of $|\Gamma'(0,\Theta)^2-\Gamma'(\pi/2,\Theta)^2|$
exists for some angle $\Theta_M$ on the Fermi surface for fixed
spin-orbit interactions.

First, we show that when $\Gamma'(0,\Theta)$ reaches the maximum
value for some $\Theta_M$, $\Gamma'(\pi/2,\Theta)$ must reach the
minimum at $\Theta_M$ and vice versa. From Eq. (\ref{GammaArb}),
the condition $(d/d\Theta)\Gamma'(0,\Theta)^2=0$ at some
$\Theta_M$ gives
\begin{equation}\label{ThetaM}
\tan(2\Theta_M)=\frac{2(\beta_{xx}\beta_{xy}+\beta_{yx}\beta_{yy})}{\Gamma(0)^2-\Gamma(\pi/2)^2},
\end{equation}
where $\Gamma(0)=\Gamma'(0,0)$ and
$\Gamma(\pi/2)=\Gamma'(\pi/2,0)$. It can also be shown that
$(d/d\Theta)\Gamma'(\pi/2,\Theta)|_{\Theta=\Theta_M}=0$. On the
other hand, we redefine the parameters $A$, $B$ and $C$ as
$A=\Gamma(0)^2$, $B=\Gamma(\pi/2)^2$, and
$C=\beta_{xx}\beta_{xy}+\beta_{yx}\beta_{yy}$. The second
derivative gives
$(d/d\Theta)^2\Gamma'(0,\Theta)^2=2(B-A)\cos(2\Theta)-4C\sin(2\Theta)$
and
$(d/d\Theta)^2\Gamma'(\pi/2,\Theta)^2=2(A-B)\cos(2\Theta)+4C\sin(2\Theta)=-(d/d\Theta)^2\Gamma'(\pi/2,\Theta)^2$.
Because the second derivative are opposite in sign for $\phi'=0$
and $\phi'=\pi/2$, this implies that when $\Gamma'(0,\Theta)^2$
has the maximum (minimum) value, $\Gamma'(\pi/2,\Theta)^2$ has the
the minimum (maximum) value. In conclusion,
$|\Gamma'(0,\Theta)^2-\Gamma'(\pi/2,\Theta)^2|$ must be the
maximum value when $\Theta=\Theta_M$.

The energy dispersion of the
$\widetilde{\beta}_R+\widetilde{\beta}_1$ system is of the form
$\Gamma(\phi)^2=A^2\cos^2\phi+B^2\sin^2\phi$. Interestingly, it
can be shown that for $\Theta=\Theta_M$, the energy dispersion
$\Gamma(\phi,\Theta)$ in general has the same form as that of the
$\widetilde{\beta}_R+\widetilde{\beta}_1$ system. Because we
require
$\Gamma'(\phi',\Theta)^2=(\beta_{x'x'}^2+\beta_{y'x'}^2)\cos^2\phi'+(\beta_{x'y'}^2+\beta_{y'y'}^2)\sin^2\phi'+(\beta_{x'x'}\beta_{x'y'}+\beta_{y'x'}\beta_{y'y'})\sin(2\phi')$
to have the form
$\Gamma'(\phi',\Theta)^2=A^2\cos^2\phi'+B^2\sin^2\phi'$, the
coefficient $\beta_{x'x'}\beta_{x'y'}+\beta_{y'x'}\beta_{y'y'}$
must be zero at some $\Theta$, and it is obtained from the
equation:
$(\beta_{x'x'}\beta_{x'y'}+\beta_{y'x'}\beta_{y'y'})=\cos(2\Theta)(\beta_{xx}\beta_{xy}+\beta_{yy}\beta_{yx})+(\Gamma(\pi/2)^2-\Gamma(0)^2)\sin(2\Theta)/2=0$.
This indeed gives $\Theta=\Theta_M$.

\begin{figure}[!htb]
\begin{center}
\includegraphics[width=8.4cm,height=4.2cm]{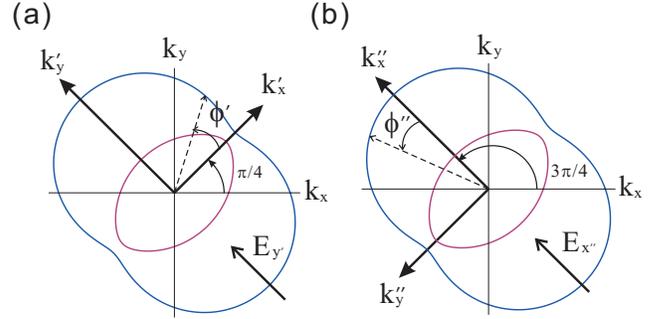}
\end{center}
\caption{(Color online) The figure shows the energy dispersion in
the Rashba-Dresselhaus system. The external electric field is
applied along the (a) $k_y'$ direction ($\Theta=\pi/4$) and (b)
$k_x''$ direction ($\Theta=3\pi/4$).}\label{figRD}
\end{figure}

Consider the Rashba-Dresselhaus system
($\widetilde{\beta}_R+\widetilde{\beta}_S$). If $k_x$ and $k_y$
lie respectively along the $[100]$ and $[010]$ directions, then
$\Gamma(0)=\Gamma(\pi/2)$ and
$\beta_{xx}\beta_{xy}+\beta_{yy}\beta_{yx}=2\alpha\beta\neq0$.
Equation (\ref{ThetaM}) implies that $\Theta_M=\pi/4,3\pi/4$. (see
\figref{figRD}) For $\Theta_M=\pi/4$, the resulting dispersion in
the new coordinate system is given by
$\Gamma'(\phi')^2=(\alpha-\beta)^2\cos^2\phi'+(\alpha+\beta)^2\sin^2\phi'$
and we have $\Gamma'(0)<\Gamma'(\pi/2)$. As a result, in order to
obtain a large spin-Hall current ($\sigma_s>e/8\pi$), an electric
field must be applied along the $k_y'$ direction because the
nearly degenerate area is located at $\Gamma'(0)$. For
$\Theta_M=3\pi/4$, we have
$\Gamma''(\phi'')^2=(\alpha+\beta)^2\cos^2\phi''+(\alpha-\beta)^2\sin^2\phi''$,
and in this case, $\Gamma''(\pi/2)<\Gamma''(0)$. The electric
field must be applied along the $k_x''$ direction for obtaining a
large spin-Hall current. As shown in \figref{figRD}, $k_x''$ is
obtained as the rotation of $k_x'$ by $\pi/2$, and thus, it is
parallel to $k_y'$.

In the Rashba-Dresselhaus system, $[110]$ and $[\bar{1}10]$ are
nonequivalent axes. The corresponding band-splitting values are
$2m(\alpha-\beta)/\hbar^2$ and $2m(\alpha+\beta)/\hbar^2$. We
change the coordinate ($k_x$, $k_y$) to ($k_x'$, $k_y'$) such that
$k_x'$ and $k_y'$ are parallel to $[110]$ and $[\bar{1}10]$,
respectively. In this case, we have $\Theta_M=\pi/4$. The
resulting effective spin-orbit matrix is
\begin{equation}\label{RDasym}
\frac{1}{\sqrt{2}}(\alpha-\beta)\left(\begin{array}{cc}
1&0\\
-1&0
\end{array}\right)+\frac{1}{\sqrt{2}}(\alpha+\beta)\left(\begin{array}{cc}
0&1\\
0&1
\end{array}\right).
\end{equation}
We have $\Gamma'(0)^2=(\alpha-\beta)^2$ and
$\Gamma'(\pi/2)=(\alpha+\beta)^2$. The asymmetric response
function $\Deltabeta'$ corresponding to the spin-orbit matrix in
Eq. (\ref{RDasym}) is $\Deltabeta'=-4\alpha\beta<0$. Using Eq.
(\ref{SHC-newc2}), we can show that
\begin{equation}\label{App:lambda}
\mathrm{Re}(\lambda'_</\lambda'_>)=-2\alpha\beta/[\alpha^2+\beta^2+|\alpha^2-\beta^2|].
\end{equation}
For $\alpha^2>\beta^2$, the ISHCs are
$\sigma^z_{x'y'}=(e/8\pi)(1+\beta/\alpha)$ and
$\sigma^z_{y'x'}=-(e/8\pi)(1-\beta/\alpha)$. For
$\alpha^2<\beta^2$, the ISHCs are
$\sigma^z_{x'y'}=-(e/8\pi)(1+\alpha/\beta)$ and
$\sigma^z_{y'x'}=(e/8\pi)(1-\alpha/\beta)$.~\cite{TWChen09RDasym}

Let us suppose that $\alpha>0$ and $\beta>0$, and in this case, we
have $\Gamma'(0)<\Gamma'(\pi/2)$. The Rashba-Dresselhaus system
has a smaller band-splitting of $2m(\alpha-\beta)/\hbar^2$ along
the $[110]$ direction on the Fermi surface. On the other hand, the
system has a larger band-splitting [$2m(\alpha+\beta)/\hbar^2$]
along the $[\bar{1}10]$ direction. When the electric field is
applied along the $k_y'$ direction ($[\bar{1}10]$), the spin-Hall
response along the $k_x'$ direction indeed has a value larger than
$e/8\pi$, i.e., $|\sigma^z_{x'y'}|>e/8\pi$, as shown above.
Interestingly, when $\alpha$ is very close to $\beta$ in
magnitude, Eq. (\ref{App:lambda}) is very close to unity. The ISHC
in the Rashba-Dresselhaus system would transit from
$\sigma^z_{x'y'}\sim+e/4\pi$ to $\sigma^z_{x'y'}\sim-e/4\pi$ upon
tuning the Rashba coupling via gate voltage.\cite{Nitta97}

Rashba coupling and Dresselhaus coupling are usually of the same
order of magnitude in the GaAs quantum well.~\cite{Rashba05RD} In
the II-VI semiconductor, Rashba coupling is larger than
Dresselhaus coupling, while in the III-V semiconductor,
Dresselhaus coupling would be larger than Rashba
coupling.~\cite{Rashba05RD} In the narrow-gap compounds, Rashba
coupling dominates.~\cite{RDmagnitude}


\begin{thebibliography}{99}

\bibitem{pri98} G. A. Prinz, Science {\bf 282}, 1660 (1998); S. A. Wolf, D. D. Awschalom,
R. A. Buhrman, J. M. Daughton, S. von Molnar, M. L. Roukes, A.Y.
Chtchelkanova, and D. M. Treger, Science {\bf 294}, 1488 (2001).
\bibitem{zut04} I. Zutic, J. Fabian, and S. D. Sarma, Rev. Mod. Phys. {\bf 76}, 323 (2004).
\bibitem{Hisch99} J. E. Hirsch, Phys. Rev. Lett. {\bf 83}, 1834
(1999).


\bibitem{Murakami03} S. Murakami, N. Nagaosa, and S.-C. Zhang, Science {\bf 301}, 1348
(2003); S. Murakami, N. Nagaosa, S.-C. Zhang, Phys. Rev. B {\bf
69}, 235206 (2004).
\bibitem{Sinova04} J. Sinova, D. Culcer, Q. Niu, N. A. Sinitsyn, T. Jungwirth, and A. H. MacDonald, Phys. Rev. Lett. {\bf 92}, 126603 (2004).
\bibitem{Kato04}  Y. K. Kato, R. C. Myers, A. C. Gossard, and D. D. Awschalom, Science {\bf 306}, 1910 (2004).
\bibitem{Stern06} N. P. Stern, S. Ghosh, G. Xiang, M. Zhu, N.
Samarth and D. D. Awschalom, Phys. Rev. Lett. {\bf 97}, 126603
(2006).
\bibitem{Sih05} V. Sih, R. C. Myers, Y. K. Kato, W. H. Lau, A. C.
Gossard, and D. D. Awschalom, Nature Physics {\bf 1}, 31 (2005).
\bibitem{Engel05} H. Engel, E. I. Rashba and B. I. Halperin, Phys.
Rev. Lett. {\bf 95}, 166605 (2005).
\bibitem{Wund05} J. Wunderlich, et al, Phys. Rev. Lett. {\bf 94},
047207 (2005).
\bibitem{Berne05} B. A. Bernevig and S.-C. Zhang, Phys. Rev. Lett.
{\bf 95}, 016801 (2005).
\bibitem{Chang06} H. J. Chang, T.-W. Chen, J. W. Chen, W. C. Hong, W. C. Tsai, Y. F. Chen and G. Y. Guo, Phys. Rev. Lett. {\bf 98}, 136403
(2006).
\bibitem{Saitoh06} E. Saitoh, M. Ueda, H. Miyajima and G. Tatara,
Appl. Phys. Lett. {\bf 88}, 182509 (2006); S. O. Valenzuela and M.
Tinkham, Nature (London) {\bf 442}, 176 (2006); K. Ando, Y.
Kajiwara, S. Takahashi, S. Maekawa, K. Takemoto, M. Takatsu and E.
Saitoh, Phys. Rev. B {\bf 78}, 014413 (2008).
\bibitem{Kimura07}T. Kimura, Y. Otani, T. Sato, S. Takahashi, and S. Maekawa, Phys. Rev. Lett. {\bf 98}, 156601 (2007).
\bibitem{Guo08} G. Y. Guo, S. Murakami, T.-W. Chen and Nagaosa,
Phys. Rev. Lett. {\bf 100}, 096401 (2008).
\bibitem{boh03} A. Bohm, A. Mostafazadeh, H. Koizumi, Q. Niu and J. Zwanziger,
{\it The Geometric Phase in Quantum Systems} (Springer, Berlin,
2003).
\bibitem{Seki08} T. Seki, Y. Hasegawa, S. Mitani, S. Takahashi, H.
Imamura, S. Maekawa, J. Nitta and K. Takahashi, Nature Mater. {\bf
7}, 125 (2008).
\bibitem{RD} S.-Q. Shen, Phys. Rev. B {\bf 70}, 081311(R)
(2004); N. A. Sinitsyn, E. M. Hankiewicz, W. Teizer, and J.
Sinova, Phys. Rev. B {\bf 70}, 081312(R) (2004); M.-C. Chang,
Phys. Rev. B {\bf 71}, 085315 (2005); T.-W. Chen, C.-M. Huang and
G. Y. Guo, Phys. Rev. B {\bf 73}, 235309 (2006).
\bibitem{BernePRB05} B. A. Bernevig and S.-C. Zhang, Phys. Rev. B
{\bf 72}, 115204 (2005); Y. Kato, R. C. Myers, A. C. Gossard, D.
D. Awschalom, Nature (London) {\bf 427}, 50 (2004); Y. Kato, R. C.
Myers, A. C. Gossard, D. D. Awschalom, Phys. Rev. Lett. {\bf 93},
176601 (2004).


\bibitem{BRashba} E. I. Rashba, Sov. Phys. Solid State {\bf 2}, 1224 (1960); Y. A. Bychkov and E. I. Rashba, J. Phys. C {\bf 17}, 6039 (1984).
\bibitem{Dre55} G. Dresselhaus, Phys. Rev. {\bf 100}, 580 (1955).
\bibitem{McClure56} J. W. McClure, Phys. Rev. {\bf 104}, 666
(1956); D. P. DiVincenzo and E. J. Mele, Phys. Rev. B {\bf 29},
1685 (1984); V. P. Gusynin and S. G. Sharapov, {\bf 95}, 146801
(2005); 19.K. S. Novoselov, A. K. Geim, S. V. Morozov, D. Jiang,
Y. Zhang, S. V. Dubonos, I. V. Grigorieva, and A. A. Firsov,
Science {\bf 306}, 666 (2004); K. S. Novoselov, A. K. Geim, S. V.
Morozov, D. Jiang, M. I. Katsnelson, I. V. Grigorieva, S. V.
Dubonos, and A. A. Firsov, Nature (London) {\bf 438}, 197 (2005);
Y. Zhang, J. P. Small, M. E. S. Amori, and P. Kim, Phys. Rev.
Lett. {\bf 94}, 176803 (2005); Y.-W. Tan, H. L. Stormer, and P.
Kim, Nature (London) {\bf 438}, 201 (2005); Y. Zhang, Y.-W. Tan,
H. L. Stormer, and P. Kim, {\bf 438}, 201 (2005).
\bibitem{Sin07} N. A. Sinitsyn, A. H. MacDonald, T. Jungwirth,
V. K. Dugaev and J. Sinova, Phys. Rev. B {\bf 75}, 045315 (2007).
\bibitem{Pikus84} G. E. Pikus and A. N. Titkov, {\it Optical
Orientation}(North-Holland, Amsterdam. 1984), p. 73.
\bibitem{kzterm} In constructing the spin-orbit matrix $\widetilde{\beta}_2$, we have
neglected the term $Dk_z(\epsilon_{yy}-\epsilon_{xx})$ which
exists when $\epsilon_{xx}\neq\epsilon_{yy}$. However, since $k_z$
is not influenced by the in-plane electric field, $\langle
k_z\rangle=0$, we can approximately neglect this term. (see
also~\cite{BernePRB05})
\bibitem{Mahan} G. D. Mahan, {\it Many-Particle Physics} (Kluwer Academic/Plenum publisher, New York,
2000).
\bibitem{Shi06} D. Culcer, J. Sinova, N. A. Sinitsyn, T. Jungwirth and A. H. MacDonald and Q. Niu, Phys. Rev. Lett. {\bf 93}, 046602 (2004); J. Shi, P. Zhang, D. Xiao and Q. Niu, Phys. Rev.
Lett. {\bf 96}, 076604 (2006).
\bibitem{BernePRB05-2} B. A. Bernevig, Phys. Rev. B {\bf 71},
073201 (2005).

\bibitem{Foldy} L. L. Foldy and S. A. Wouthuysen, Phys. Rev. {\bf
78}, 29 (1950); T.-W. Chen and D.-W. Chiou, Phys. Rev. A. {\bf
82}, 012115 (2010).
\bibitem{Manuel11} M. Valin-Rodriguez, Phys. Rev. Lett. {\bf
107}, 266801 (2011).




\bibitem{arf95} G. Arfken and H. J. Weber, {\it Mathematical Methods for Physicists}
(Academic Press, 1995).




\bibitem{TWChen09Dis} J. Sinova, S. Murakami, S.-Q. Shen and M.-S. Choi, Solid State Commun. {\bf
138}, 214 (2006); J. I. Inoue, G. E. W. Bauer and L. W. Molenkamp,
Phys. Rev. B {\bf 70}, 041303(R) (2004); E. G. Mishchenko, A. V.
Shytov and B. I. Halperin, Phys. Rev. Lett. {\bf 93}, 226602
(2004); O. Chalaev and D. Loss, Phys. Rev. B {\bf 71}, 245318
(2005); O. V. Dimitrova, Phys. Rev. B {\bf 71}, 245327 (2005).
\bibitem{Kh06} J. Schliemann and D. Loss, Phys. Rev. B {\bf 69}, 165315
(2004); A. G. Mal'shukov and K. A. Chao, Phys. Rev. B{\bf 71},
121308 (R) (2005); A. Khaetskii, Phys. Rev. B {\bf 73}, 115323
(2006); N. Sugimoto, S. Onoda. S. Murakami and N. Nagaosa, Phys.
Rev. B {\bf 73}, 113305 (2006).
\bibitem{Burkov04} A. A. Burkov, Alvaro S. Nunez and A. H. MacDonald, Phys. Rev. B {\bf 70}, 155308 (2004).
\bibitem{Nomura05} K. Nomura, J. Sinova, T. Jungwirth, Q. Niu, and A. H.
MacDonald, Phys. Rev. B {\bf 71}, 041304 (2005); B. K. Nikolic, L.
P. Zarbo, and S. Souma, Phys. Rev. B {\bf 72}, 075361 (2005); L.
Sheng, D. N. Sheng, C. S. Ting, Phys. Rev. Lett. {\bf 94}, 016602
(2005); E. M. Hankiewicz, L. W. Molenkamp, T. Jungwirth, and J.
Sinova, Phys. Rev. B {\bf 70}, 241301(R) (2004).
\bibitem{Sherman10}V. K. Duagev, M. Inglot, E. Ya. Sherman, and J.
Barna$\acute{s}$, Phys. Rev. B {\bf 82}, 121310(R) (2010).
\bibitem{Rai12} R. Raimondi, P. Schwab, C. Gorini, and G. Vignale, Ann.
Phys. (Berlin) 524, 153 (2012) [arXiv:1110.5279].
\bibitem{TWChen09RDasym}
T.-W. Chen, H.-C. Hsu and G. Y. Guo, Phys. Rev. B {\bf 80}, 165302
(2009).



\bibitem{Rashba03} E. I. Rashba, Phys. Rev. B {\bf 68},
241315(R) (2003).
\bibitem{Rashba04} E. I. Rashba, Phys. Rev. B {\bf 70}, 161201(R)
(2004); A. A. Burkov, A. S. N$\acute{u}\tilde{n}$ez and A. H.
MacDonald, Phys. Rev. B {\bf 70}, 155308 (2004); E. B. Sonin,
Phys. Rev. B {\bf 76}, 033306 (2007).
\bibitem{Sun07} Q.-f. Sun, X. C. Xie and J. Wang, Phys. Rev. Lett.
{\bf 98}, 196801 (2007); Phys. Rev. B {\bf 77}, 035327 (2008).
\bibitem{Sab08} V. A. Sablikov, A. A. Sukhanov and Y. Ya. Tkach,
Phys. Rev. B {\bf 78}, 153302 (2008).
\bibitem{Dro11} H.-J. Drouhin, G. Fishman and J.-E. Wegrowe, Phys.
Rev. B {\bf 83}, 113307 (2011).
\bibitem{Sonin07} E. B. Sonin, Phys. Rev. Lett. {\bf 99}, 266602
(2007).
\bibitem{Tokatly08} I. V. Tokatly, Phys. Rev. Lett. {\bf 101},
106601 (2008).




\bibitem{Nitta97} J. Nitta, T. Akazaki, H. Takayanagi and T.
Enoki, Phys. Rev. Lett. {\bf 78}, 1335 (1997).
\bibitem{Rashba05RD} E. I. Rashba, Physica E: Low-dimensional Systems and Nanostructures
{\bf 34}, 31 (2006); Hyun C. Lee and S.-R. Eric Yang, Phys. Rev. B
{\bf 72}, 245338 (2005).

\bibitem{RDmagnitude} B. Jusserand, D. Richards, G. Allan, C. Priester and B. Etienne, Phys. Rev. B {\bf 51}, 4707 (1995);
W. Knap, C. Skierbiszewski, A. Zduniak, E. Litwin-Staszewska, D.
Bertho, F. Kobbi, J. L. Robert, G. E. Pikus, F. G. Pikus, S. V.
Iordanskii, V. Mosser, K. Zekentes, and Yu. B. Lyanda-Geller,
Phys. Rev. B {\bf 53}, 3912 (1996); J. B. Miller, D. M. Zumbuhl,
C. M. Marcus, Y. B. Lyanda-Geller, D. Goldhaber-Gordon, K.
Campman, and A. C. Gossard, Phys. Rev. Lett. {\bf 90}, 076807
(2003).





\end{thebibliography}
\end{document}